\documentclass[twocolumn]{aastex701}

\begin{document}

\title{No Surviving Companion to the Galactic SN 1181: \\
Evidence for a Double-Degenerate Channel for Type Iax Supernovae}

\author[orcid=0000-0002-6765-8988]{Kohki Uno}
\affiliation{Department of Astronomy and Columbia Astrophysics Laboratory, Columbia University, New York, NY, USA}
\email[show]{ku2204@columbia.edu}  

\author[orcid=0000-0002-6347-3089]{Daichi Tsuna}
\affiliation{Center for Astrophysics $|$ Harvard \& Smithsonian, 60 Garden St, Cambridge, MA 02138, USA}
\email{daichi.tsuna@cfa.harvard.edu}  

\author[orcid=0000-0002-1125-9187]{Daichi Hiramatsu}
\affiliation{Department of Astronomy, University of Florida, Bryant Space Science Center, Gainesville, FL 32611-2055, USA}
\email{dhiramatsu@ufl.edu}  

\author[orcid=0000-0002-3033-4576]{Tomoya Kinugawa}
\affiliation{Faculty of Engineering, Shinshu University, 4-17-1, Wakasato, Nagano-shi, Nagano, 380-8553, Japan}
\affiliation{Research Center for Advanced Air-mobility Systems, Shinshu University, 4-17-1, Wakasato, Nagano-shi, Nagano, 380-8553, Japan}
\affiliation{Research Center for the Early Universe (RESCEU), School of Science, The University of Tokyo, Bunkyo, Tokyo 113-0033, Japan}
\email{kinugawa@shinshu-u.ac.jp}  

\author[orcid=0000-0001-6842-5441]{Takatoshi Ko}
\affiliation{Research Center for the Early Universe (RESCEU), School of Science, The University of Tokyo, Bunkyo, Tokyo 113-0033, Japan}
\affiliation{Department of Astronomy, School of Science, The University of Tokyo, Bunkyo, Tokyo 113-0033, Japan}
\email{ko.takatoshi@gmail.com}  

\begin{abstract}

Type Iax supernovae (SNe Iax) are the recently-established, yet peculiar subclass of thermonuclear supernovae, whose progenitor systems and explosion mechanisms remain debated. A leading scenario is a single-degenerate channel, in which a white dwarf accreting from a He-star companion undergoes a pure deflagration, leaving both a bound remnant and surviving companion. The Galactic SN 1181, whose distinctive properties are consistent with an SN Iax involving a weak explosion and bound white dwarf remnant, offers a unique opportunity to test this scenario. Here, we perform a deep search for a surviving He-star companion, a hallmark of the single-degenerate He-donor channel, within $30^{\prime\prime}$ ($\sim 0.3$ pc) of the remnant using archival Gaia and Pan-STARRS1 data. The parallaxes and proper motions exclude all Gaia sources, while the spectral energy distributions of the remaining Pan-STARRS1 sources are inconsistent with both the He-star spectral templates and known hot-subdwarf population. Binary evolution calculations predict a minimum mass of a companion with an absolute $g$-band magnitude of $M_{g} \lesssim 6.5$ mag, much brighter than the Pan-STARRS1 detection limit of $M_{g} > 8$ mag. Our non-detection rules out He-star and luminous hydrogen-rich companions for SN 1181, favoring a double-degenerate channel (i.e., a white dwarf merger). Adding to the luminous He-star companion identified in the pre-explosion imaging of SN 2012Z, our results provide direct evidence for multiple progenitor channels leading to SNe Iax.

\end{abstract}

\keywords{\uat{Type Ia supernova}{1728} --- \uat{Supernova remnants}{1667} --- \uat{White dwarfs}{1799} --- \uat{Stellar evolution }{1599}}

\section{Introduction} 

Type Ia supernovae (SNe Ia) have been understood as thermonuclear explosions of CO white dwarfs \citep[WDs;][]{Whelan1973ApJ,Nomoto1982ApJ,Webbink1984ApJ, Iben1984ApJS}; however, the progenitor systems and explosion mechanisms of SNe Ia remain actively debated. Two leading progenitor channels are: the single-degenerate (SD) scenario \citep{Whelan1973ApJ,Nomoto1982ApJ}, in which a WD accretes from a non-degenerate companion, and the double-degenerate (DD) scenario \citep{Webbink1984ApJ, Iben1984ApJS}, involving the merger of two WDs. Likewise, several explosion mechanisms have been proposed. In the delayed-detonation model \citep[e.g.,][]{Maeda2010ApJ, Seitenzahl_2013MNRAS.429.1156S}, a subsonic deflagration ignited in the CO core transitions into a supersonic detonation via the deflagration-to-detonation transition (DDT), while in the double-detonation model \citep[e.g.,][]{Shen_2009ApJ...699.1365S, Fink_2010AA...514A..53F,Pakmor_2013ApJ...770L...8P}, the first detonation in the He-shell drives a converging shock that triggers a second detonation in the CO core \citep[see e.g.,][for reviews]{Maoz2014Ia,Maeda2022hxga,Ruiter25}. 

Recent transient surveys have revealed a remarkable diversity among SN Ia-like events \citep{Taubenberger_2017hsn..book..317T}, indicating that SNe Ia comprise multiple populations arising from different progenitor systems and explosion mechanisms. Among these subclasses, Type Iax SNe \citep[SNe Iax;][]{Foley_2013ApJ...767...57F,Jha_2017hsn..book..375J}, or SN 2002cx-like events \citep{Li_2003PASP..115..453L}, are characterized by lower peak luminosities ($M_{V}\approx -13$ to $-19$ mag) and lower ejecta velocities ($\sim 2000-8000$ km s$^{-1}$) than canonical SNe Ia. They contribute $\sim 10-30 \%$ of the SN Ia rate \citep{Foley_2013ApJ...767...57F, Srivastav_2022MNRAS.511.2708S}. 

A particularly suggestive clue to their progenitor systems comes from pre- and/or post-explosion imaging of nearby events, in which detected point sources at the SN position have been interpreted as luminous companion stars or bound remnants. The relatively bright SN Iax 2012Z \citep{McCully14,McCully_2022ApJ...925..138M}, for example, showed a luminous blue source in pre-explosion imaging consistent with a He-star companion. These observations have motivated the pure(failed)-deflagration scenario \citep[e.g.,][]{Jordan_2012ApJ...761L..23J,Kromer_2013MNRAS.429.2287K,Fink14}, in which He-rich accretion onto a WD ignites a deflagration that fails to undergo DDT, partially ejecting the WD and leaving a bound remnant. Typically discussed within the SD-like framework, this scenario reasonably reproduces the photometric and spectroscopic properties of bright SNe Iax, but systematically fails to account for the faint members of SNe Iax, such as SN 2008ha \citep[][but see also \citealt{2015MNRAS.450.3045K}]{Foley_2013ApJ...767...57F} and 2021fcg \citep{Karambelkar_2021ApJ...921L...6K}. Since faint SNe Iax may constitute a substantial fraction of the overall SN Iax rate, clarifying the physical origin of the bright and faint populations is essential for a comprehensive picture of the entire SN Iax population, and more broadly for the role of failed WD explosions.

A unique opportunity to probe the faint end of the SN Iax population is offered by the Galactic SN 1181. Historical records relate the SN to Saturn, leading to an apparent magnitude estimate of $m_V \approx -1.4$ to $1.0$ mag \citep{Ritter21,Schaefer23}. Combined with the extinction and the distance of $D \approx 2.3$--$2.5$ kpc \citep{Bailer-Jones21,Lykou23}, this implies a peak absolute magnitude of $M_V \approx -13$ to $-16$ mag, placing SN 1181 within the faint SN Iax regime (see e.g., Figure 4 of \citealt{Srivastav_2022MNRAS.511.2708S}). Moreover, a peculiar hot WD, IRAS 00500+6713, is located at a position consistent with the historical records. The WD is characterized by its very high temperature, extremely fast wind, and O/Ne/Mg-rich abundance \citep{Gvaramadze19,Oskinova20,Lykou23}. These properties are broadly consistent with a partially burned, bound remnant of a failed thermonuclear explosion like SN Iax.

If SN 1181 shares a common progenitor channel with bright SNe Iax, namely an SD-like pure-deflagration explosion driven by He accretion, the donor He star should have survived and remain detectable around the remnant. On the other hand, the absence of such a companion would disfavor the SD-like He-donor channel for SN 1181, suggesting that the faint SN Iax population follows a different evolutionary path from bright ones. Therefore, a targeted search for a surviving He-star companion around SN 1181 provides a critical test for the progenitor channel for SNe Iax. A surviving binary system has been strongly disfavored by previous searches for periodic photometric variations \citep{Schaefer23,Ko26}. However, it is still possible that the companion has been unbound upon the SN, which could be detectable by observations surrounding the remnant WD.

In this Letter, we conduct a search for a surviving He-star companion around the remnant of SN 1181. In Section~\ref{sec:companion_search}, we investigate the archival Gaia and Pan-STARRS data around SN 1181, finding no plausible candidates consistent with the expected properties of the He stars. Then in Section \ref{sec:binary_models}, we perform binary stellar evolution calculations, excluding the possibility of a He star-WD binary as the progenitor system. Finally in Section~\ref{sec:conclusions}, we discuss the implications of this non-detection for the progenitor system of SN 1181 and for the evolutionary pathways leading to faint SNe Iax.

\section{Helium Star Companion Search} \label{sec:companion_search}

\begin{figure*}[htb]
\centering
\epsscale{1.17}
\plotone{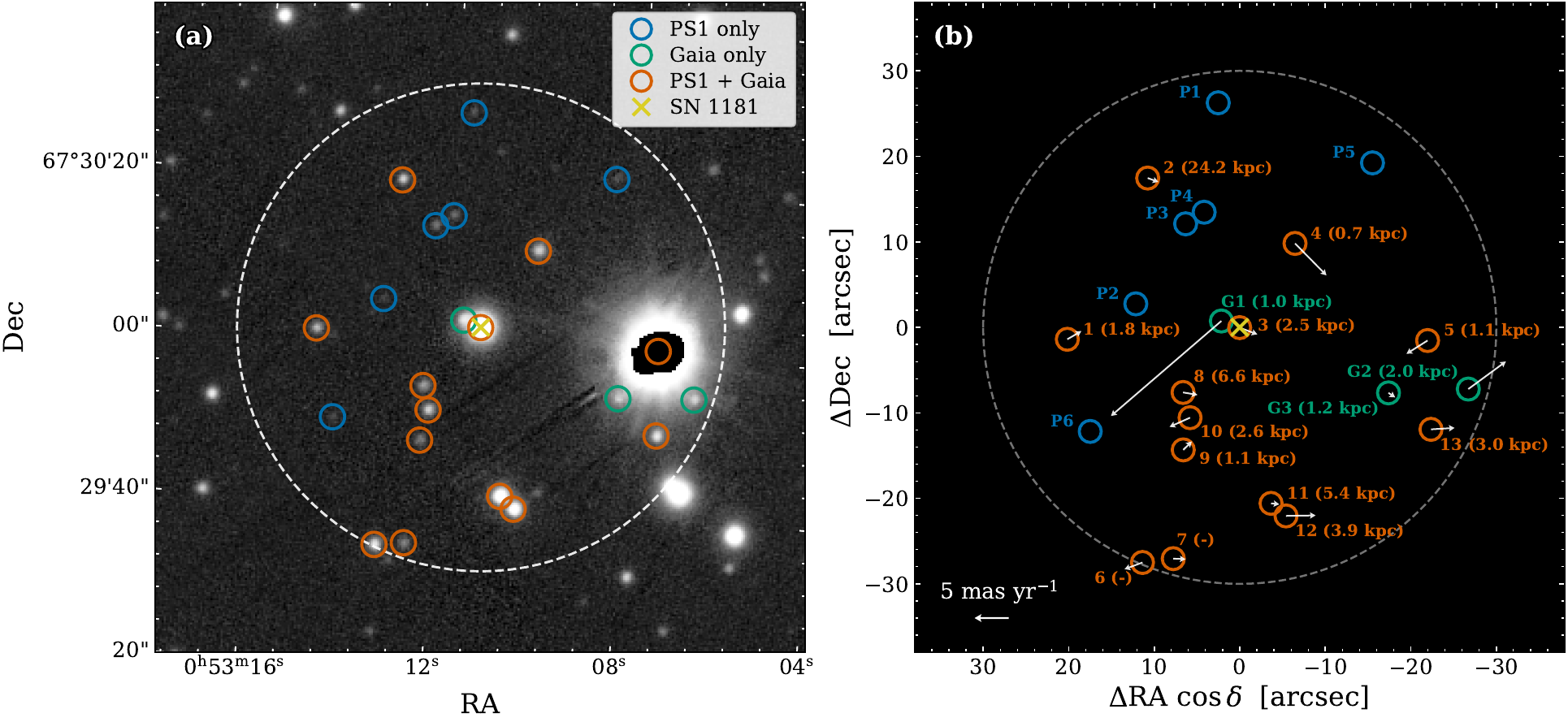}
\caption{
(a) Pan-STARRS1 $i$-band image with the field-star identification. The dashed white circle marks the $30^{\prime\prime}$ search radius around the central remnant of SN 1181 (yellow cross). The identified sources are circled and color-coded by cross-match class: Pan-STARRS1-only (blue), Gaia-only (green), and both (orange). 
(b) Relative source positions in the tangential plane centered on SN 1181. The sources are color-coded as in (a) and labeled by source ID (P1–P6 for PS1-only, G1-G3 for Gaia-only, and $n$ for both), with their distance estimates from Gaia. White arrows show the Gaia proper-motion vectors.
}
\label{fig:crossmatch}
\end{figure*}

We search for a surviving companion within $30^{\prime\prime}$ of SN 1181 using the data from Gaia \citep{Gaia_2016AA...595A...1G,Gaia_2023AA...674A...1G} and Pan-STARRS1 \citep[PS1;][]{Kaiser_2002SPIE.4836..154K, Chambers_2016arXiv161205560C}. At the distance of SN 1181 (which we adopt $2.5$\,kpc in this work), this angular radius corresponds to a projected physical separation of $\sim 0.3$ pc, which is sufficiently large to encompass a surviving He-star companion ejected by the supernova explosion. The remnant WD has a projected velocity of $\sim 30$ km s$^{-1}$ \citep{Lykou23}, while the companion could have received a kick of up to a few $\times 100$ km s$^{-1}$, depending on the orbital configuration and projection effects. Thus, the companion could have traveled a projected distance comparable to the adopted search radius since the explosion. Within this region, we identified 16 Gaia sources and 19 PS1 sources (Figure~\ref{fig:crossmatch}): 13 sources in both surveys, 3 only in Gaia, and 6 only in PS1, resulting in a total of 22 unique sources.

\subsection{Gaia Parallax and Proper-Motion Constraints}

For the sources detected by Gaia, the parallax and proper-motion measurements allow us to assess their distances and projected trajectories, respectively. A surviving companion ejected in the SN 1181 explosion should lie at a distance consistent with that of the remnant ($2.5$ kpc), and relative to the central WD, should trace back to the explosion site. We reject a Gaia source if its parallax is inconsistent with the adopted distance of SN 1181 or if its proper-motion trajectory does not intersect the allowed explosion-site region (within a radius of $3^{\prime\prime}$ around the WD expected from its projected velocity of $\sim 30$ km s$^{-1}$ since the explosion). No Gaia source satisfies both criteria, and we therefore find no plausible Gaia-detected candidate for a surviving companion associated with the progenitor system of SN 1181. This leaves the six PS1-only sources as the remaining cataloged candidates, which we examine using their optical photometric properties in the following.

\subsection{PS1 Spectral Energy Distribution Constraints}

\begin{figure*}[htb]
\centering
\epsscale{1.17}
\plotone{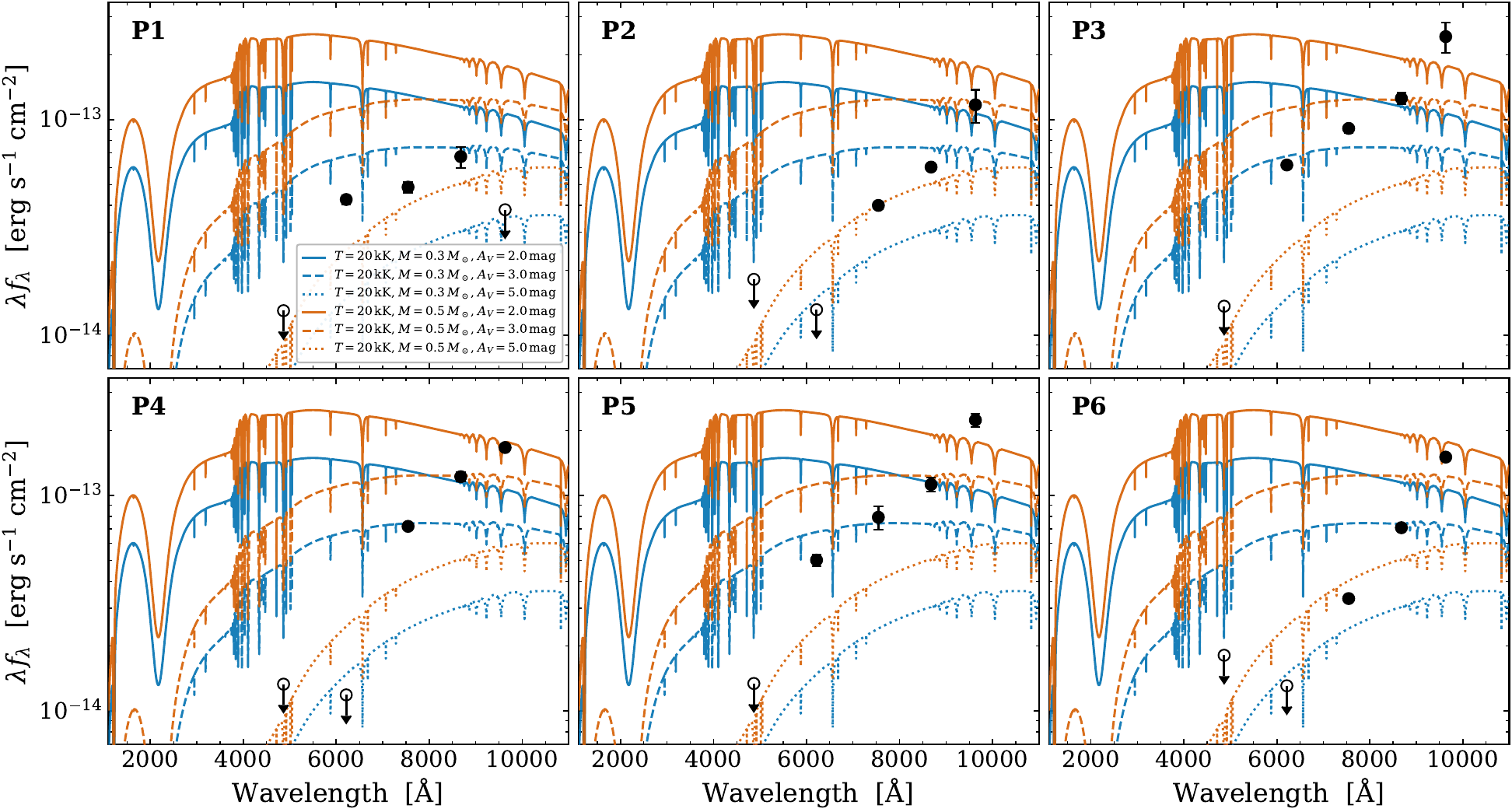}
\caption{
SEDs of the six companion candidates (P1–P6) with the He-star spectral models. Black filled circles are PS1 $grizy$ PSF magnitudes with 1$\sigma$ error bars, while open circles with downward arrows denote upper limits. Background spectra are the He-star atmosphere models with different masses, temperatures, and extinctions calculated by the extinction law of \citet{Fitzpatrick_1999_PASP_111_63F} using $R_{V}=3.1$. Their fluxes are scaled to the adopted distance to SN 1181 of 2.5 kpc.
}
\label{fig:sed}
\end{figure*}

Figure \ref{fig:sed} shows the spectral energy distributions (SEDs) of the six possible surviving companions detected only in PS1. All six sources exhibit red optical SEDs, which may arise either from substantial interstellar extinction toward SN 1181, located close to the Galactic plane, or from intrinsically low effective temperatures. In the latter case, the sources would be ordinary cool stars and could not be He-star companions, which are expected to have effective temperatures of at least $10^{4}$ K (see Figure \ref{fig:Hestar_mag_hist} for their temperature distribution). Therefore, we examined whether the observed red SEDs could instead be explained by intrinsically hot He stars with strong interstellar extinction.

We compare the PS1 SEDs with He-star spectral templates \citep{Rauch_2003_AA_403_709R}, applying the extinction law of \citet{Fitzpatrick_1999_PASP_111_63F} with $R_{V}=3.1$ and $A_{V}=2.0$-$3.0$ mag. The $A_{V}$ range was estimated from nearby field stars within $2^{\prime}$ \citep{Lykou23}, which is also consistent with the 3D dust map \citep{Green_2015ApJ...810...25G,Green_2018JOSS....3..695G}. For the models, we adopt a He-rich composition with hydrogen and helium mass fractions of $X_{\mathrm{H}}=0.2$ and $X_{\mathrm{He}}=0.8$, respectively, an effective temperature of $T_{\mathrm{eff}}=2\times10^{4}$ K, and a surface gravity of $\log g=6$. These parameter choices are conservative because they produce relatively red optical colors and, together with the low stellar masses considered below, minimize the predicted optical flux within the plausible parameter range. For each assumed He-star mass, we calculate the stellar radius as $R_{\mathrm{He}}=\sqrt{GM_{\mathrm{He}}/g}$ and scale the model flux to the adopted distance of SN 1181. We consider $M_{\mathrm{He}}=0.3$-$0.5,M_{\odot}$, covering the lowest plausible mass range for a surviving He-star companion in the evolutionary scenario considered here (see Section~\ref{sec:binary_models} for details).

Among the models considered, the observed red SEDs are most closely approximated by strongly attenuated He stars at the lower end of this mass range. However, even the faintest and most strongly extinguished models cannot simultaneously reproduce the observations, particularly the $g$-band upper limits and the red optical colors, such as $i-z$. We therefore conclude that none of the six PS1-only sources is consistent with the He-star models considered here.

We next examine whether a surviving He star could have escaped detection in the PS1 data. The mean $5\sigma$ point-source limiting magnitudes of PS1\footnote{$(g,r,i,z,y)=(23.3, 23.2, 23.1, 22.3, 21.4)$ mag \citep{Chambers_2016arXiv161205560C}} indicate that the He-star models within the adopted mass and extinction ranges would be detectable in at least one PS1 band. Even in the most conservative case considered here, corresponding to a $0.3,M_{\odot}$ He star with $T_{\mathrm{eff}}=2\times10^{4}$ K, we require an unrealistically large $A_{V}>5.0$ mag for the source to remain undetected in the $g$ band. However, its predicted fluxes in the redder bands, particularly $r$ and/or $i$, would exceed the corresponding detection limits. Therefore, it is unlikely that a surviving He star within the parameter range considered here escapes detection in all PS1 bands.

\begin{figure}[tb]
\centering
\epsscale{1.17}
\plotone{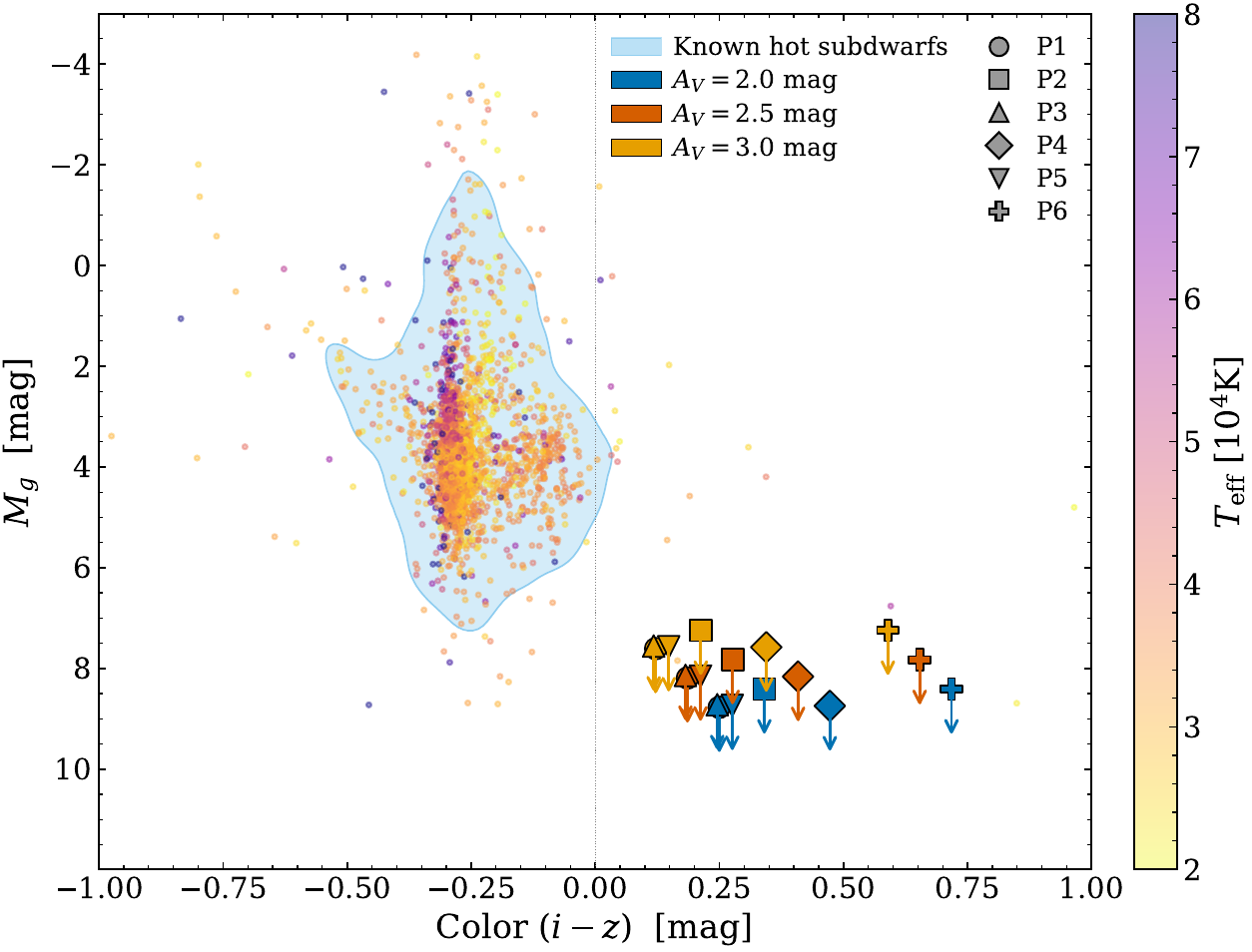}
\caption{
Color–magnitude diagram of the extinction-corrected absolute $g$-band magnitude (at the distance of SN 1181) and $i-z$ color for the six companion candidates P1–P6 (Figure \ref{fig:sed}), compared with the known hot-subdwarf population \citep{Culpan_2022_AA_662A_40C}. The scattered points are hot subdwarfs (color-coded by their effective temperatures), and the blue shaded region shows their 95\% density contour. The candidates (distinct markers) are shown for the adopted extinction values, while the arrows show $g$-band upper limits.
}
\label{fig:colors}
\end{figure}

The spectral-template comparison alone, however, may not fully exclude He-rich stars with atmospheric properties beyond the range covered by the adopted models. To test the result independently of the adopted atmosphere grid, we also perform an empirical comparison with the cataloged hot subdwarfs, which are compact, hot, stripped stars including faint He-rich objects \citep{Culpan_2022_AA_662A_40C}. Figure \ref{fig:colors} compares the characteristic photometric properties of the candidates: the $g$-band absolute magnitude and the red optical color $i-z$. Assuming that the PS1-only sources lie at the distance of SN 1181, we calculate their extinction-corrected absolute magnitudes. In the color-magnitude diagram, the six PS1-only sources occupy a region that is clearly separated from the population of the hot subdwarfs. Moreover, none of the hot-subdwarf outliers lying outside the main population simultaneously satisfies the observational constraints of the PS1-only sources. Thus, neither the standard He-star models nor the observed population of hot subdwarfs can account for the six remaining candidates.

Taken together, the Gaia astrometry, the PS1 SEDs, and the PS1 detection limits reveal no viable surviving He-star companion within the search region. Although we cannot strictly exclude an unknown class of He star with properties unlike those of both the adopted atmosphere models and currently known hot subdwarfs, our investigation provides strong evidence against the presence of a surviving He-star companion to SN 1181.

\section{Comparison to Binary Evolution Models} \label{sec:binary_models}

\begin{figure}[htb]
\centering
\includegraphics[width=\linewidth]{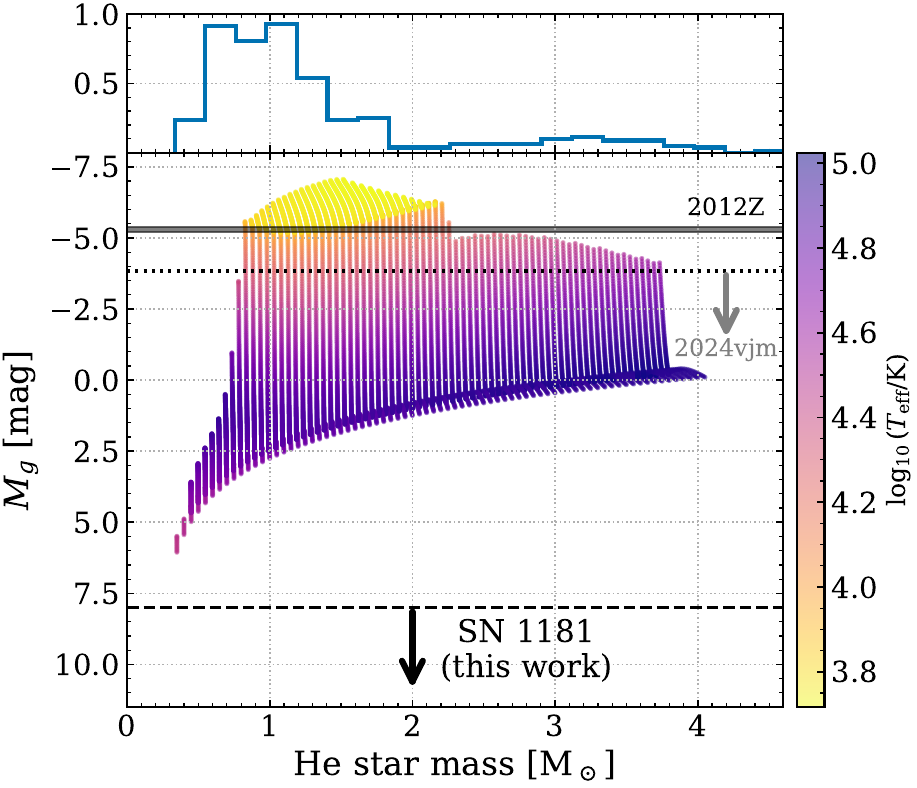}
\caption{Comparison of our $g$-band magnitude limit to He star companions of WDs predicted from binary evolution models. Top panel shows the histogram of the predicted He star mass, which shows a dearth of stars below $\approx 0.3~M_\odot$. We also show the measured luminosity of the He star companion of SN 2012Z \citep{McCully14}, and a deeper limit for SN 2024vjm \citep{Zimmerman26} that rules out an SN 2012Z-like companion. Compared to these SNe, our limit from SN 1181 (dashed line) is much deeper thanks to its proximity.}
\label{fig:Hestar_mag_hist}
\end{figure}

Searches for stellar companions to SNe Iax have been performed for SNe 2012Z \citep{McCully14}, 2008ha \citep{Foley14}, 2014dt \citep{Foley15}, and 2024vjm \citep{Zimmerman26}. A luminous He star with $M_{\rm F555W}\approx -5.3$ has been identified in the pre-explosion imaging of SN 2012Z from the {\it Hubble Space Telescope} ({\it HST}), which has strongly supported the SD scenario of a WD undergoing pure deflagration \citep{Jordan_2012ApJ...761L..23J,Kromer_2013MNRAS.429.2287K,Fink14}. For SN 2008ha, a luminous red source ($M_{\rm F814W}=-5.4$ mag) was identified 4 years after the explosion, which could be an inflated WD remnant or a companion star. For SNe 2014dt and 2024vjm, non-detections from pre-explosion images resulted in upper limits of $M_{\rm F438W}>-5.0$ mag and $M_{\rm F814W}>-5.9$ mag from the {\it HST} for SN 2014dt, and $M_{\rm VIS}>-3.84\pm 0.19$ mag from {\it Euclid} for SN 2024vjm.

Thanks to the proximity of SN 1181, the much deeper limits obtained from PS1 place stringent constraints on the putative stellar companion. Our $g$-band limit of $M_g>8$ mag immediately rules out a kicked companion with luminosities similar to that observed for SN 2012Z. Our limits for SN 1181 are also more than $10$ mag deeper than the deepest limit from SN 2024vjm, which allows us to test the SD channel for SN 1181 with an unprecedented depth.

To investigate the range of possible He star companions, we construct WD--He star binary systems using the Binary Stellar Evolution (BSE) code \citep{Hurley00, Hurley02}. The primary masses are drawn from a Salpeter initial mass function in the range $1$--$10\,M_\odot$ relevant for WD primaries. For the initial conditions of the binaries, we adopt a flat mass-ratio distribution, log-uniform separations with a minimum that avoids contact and a maximum of $10^6\,R_\odot$, and a thermal eccentricity distribution. Many of the systems of our interest undergo a common-envelope (CE) phase, and we model its outcome using the $\alpha\lambda$ formalism \citep{Webbink1984ApJ,deKool90}, adopting $\alpha=3$ and calculating $\lambda$ with the default BSE prescription using a parameter value of 0.5. The $\alpha\lambda$ parameter mainly affects the post-CE separation and thus not the He star mass of our interest, and we verify that their mass range remains largely unchanged even when $\alpha\lambda$ is varied from $0.01$ to $100$ times the default value. As a conservative approach, no constraint is imposed on the orbital separation of WD--He binaries, though the majority of systems are close binaries\footnote{This leads to an orbital velocity of $v_{\rm orb}>100$ km s$^{-1}$ for a total mass of $>0.5~M_\odot$. Thus for an unbound binary, chance alignment between the WD and the He star within the angular resolution of Pan-STARRS (0.7 arcsec; corresponding to tangential velocity of 10 km s$^{-1}$) would be improbable at a $<0.25\%$ level.} with separations of $\lesssim 10\,R_\odot$.

We select systems with primary WD masses of $1.1$--$1.3\,M_\odot$, based on prior observations and modeling favoring a massive WD remnant \citep[e.g.,][]{Kashiyama19,Oskinova20,Lykou23}. The resulting companion-mass distribution, shown in the top panel of Figure \ref{fig:Hestar_mag_hist}, shows a broad peak around $0.5$--$2\,M_\odot$, and extends down to a minimum mass of $0.3\,M_\odot$. This lower limit corresponds to the minimum mass of the He core to sustain nondegenerate core He burning \citep[e.g.,][]{Han02,Kippenhahn13}, and is robust to the uncertain binary physics.

In the bottom panel of Figure \ref{fig:Hestar_mag_hist}, we show the predicted $g$-band magnitudes of the He star companions. As the SD channel involves a mass-transferring donor, we extract the evolutionary phase from helium zero-age main sequence up to maximum radii, beyond which the He stars contract (the He star models barely change from maximum radii, over the timescale of $\approx 840$ years from SN 1181). We calculate the $g$-band magnitudes from the radii and effective temperatures from BSE assuming blackbody spectra, which we find to be a good approximation in the $g$-band within $\approx 0.3$ mag when compared to more detailed SED models of He stars used in Section \ref{sec:companion_search}. The minimum mass of the He star translates to an absolute $g$-band upper limit of $M_g\lesssim 6.5$ mag\footnote{This estimate is conservative, as previous numerical works have shown that these companions could inflate and brighten after the SN due to interaction with the SN ejecta \citep[e.g.,][]{Pan13,Bauer19}.}, and our limit of $M_g>8$ mag therefore rules out a He star companion to SN 1181.

\section{Discussion and Conclusions} \label{sec:conclusions}

Our observational and numerical results strongly disfavor the presence of a surviving He-star companion for SN 1181. Before discussing the implications, below we summarize the main findings of our work:
\begin{itemize}
    \item The Gaia parallax and proper-motion measurements exclude all 16 Gaia-detected sources as plausible surviving companions. The 6 remaining PS1 sources are inconsistent with both He-star atmosphere models and the known population of hot subdwarfs.
    \item Our binary stellar evolution calculations provide a theoretical lower mass limit of $\sim 0.3 M_{\odot}$ for the He-star companion, corresponding to an absolute $g$-band magnitude of $M_{g} \lesssim 6.5$ mag. The PS1 detection limit of $M_{g}>8$ mag is sufficiently deep to detect even the faintest He-star companions predicted by these calculations.
\end{itemize}

\begin{figure}
\centering
\includegraphics[width=\linewidth]{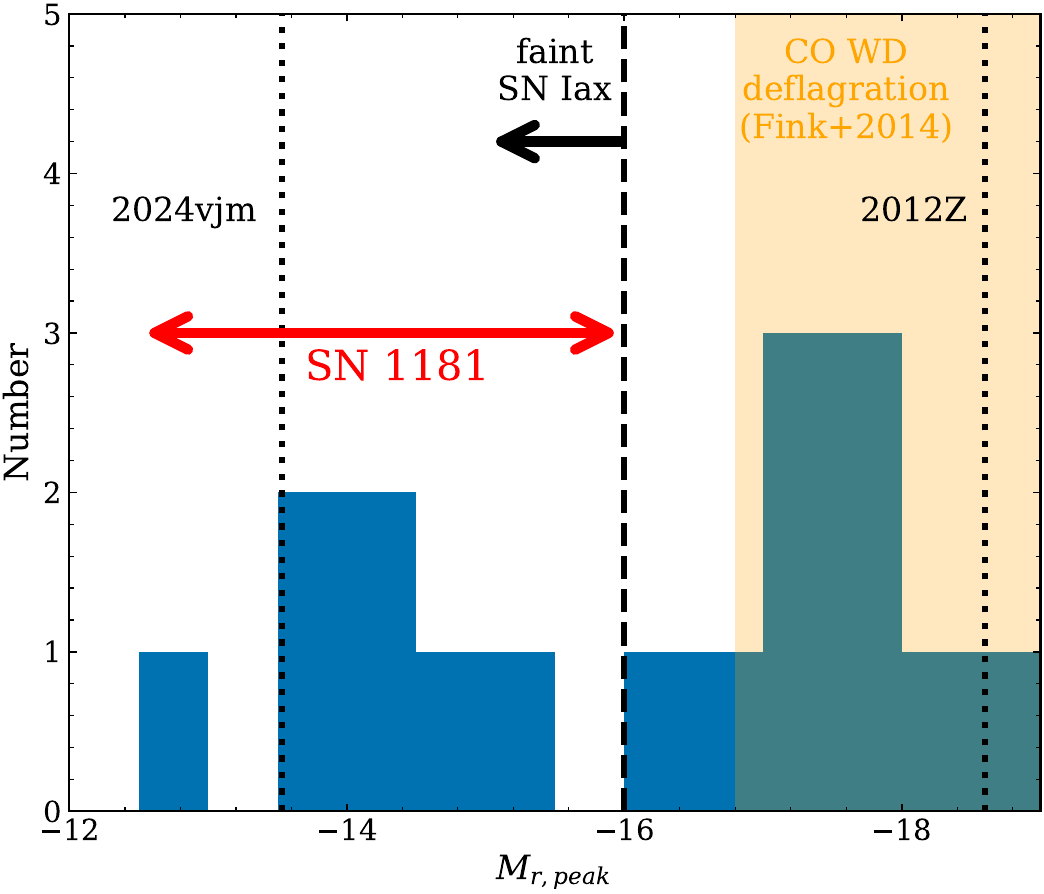}
\caption{Comparison of the peak $r$-band magnitude of SN 1181 \citep[based on historical records;][]{Ritter21,Schaefer23} and a sample of SN Iax taken from \cite{Zimmerman26}. SN 1181 belongs to the ``faint" SN Iax group (based on \citealt{Srivastav_2022MNRAS.511.2708S}), which is not well reproduced by the standard pure deflagration models of CO WDs \citep{Fink14}. Dotted lines show the peak magnitudes of SN 2012Z and SN 2024vjm.}
\label{fig:Iax_hist}
\end{figure}

\noindent Along with He-star donors, H-rich companions such as red giants have also been proposed for SN 2008ha, motivated by the detection of a luminous ($\sim -5$ mag) red source. At the distance of SN 1181, however, an analogously bright source should have been detected by Gaia. Therefore, our results disfavor not only a He-star donor but also the luminous H-rich companions previously proposed for SNe Iax. Beyond previous suggestions based on theoretical interpretation and constraints on an unresolved close companion \citep[e.g.,][]{Oskinova20, Schaefer23, Ko_2024ApJ...969..116K}, our wide-field survey across the kinematically allowed region provides the comprehensive, direct observational evidence against a non-degenerate companion in the SN 1181 progenitor system, thereby strongly favoring a double-degenerate WD-merger scenario. In this scenario, the source associated with SN 2008ha may instead be the bound remnant of the progenitor WD \citep[e.g.,][]{Shen_2009ApJ...699.1365S}.

Our conclusion for SN 1181 reflects a diversity of progenitor and explosion channels within SNe Iax. The broad luminosity distribution of SNe Iax spanning from $-13$ mag to $-18$ mag has motivated to split the SNe Iax into subpopulations: faint and luminous sequences (Figure \ref{fig:Iax_hist}). Our results reinforce multiple pathways leading to SNe Iax: luminous SN 2012Z-like events originating from pure deflagration in He star-WD systems, while faint events such as SN 1181 arising from WD mergers. The CO WD deflagration model through SD-like scenarios successfully explains the luminous SNe Iax, but not the fainter population in a single unified picture of progenitor system and explosion mechanism. Our work observationally demonstrates the need for an alternative scenario for the faint SNe Iax other than the CO WD deflagration model, and further motivates numerical simulations of WD mergers \citep[e.g.,][]{Kashyap18}.

The current observed sample of SNe Iax remains small and biased due to their intrinsically faint luminosity, precluding a comprehensive understanding of their nature. The unprecedented depths of next-generation transient surveys, such as the \textit{Nancy Grace Roman Space Telescope} \citep{Rose_2021arXiv211103081R} and the Vera C. Rubin Observatory \citep{LSST_2019ApJ...873..111I}, will substantially enlarge the SN Iax sample, particularly the fainter events. Furthermore, they will also provide a rich dataset of pre-explosion images for future nearby SNe Iax, enabling direct constraints on their progenitor systems.

\section*{acknowledgments}

K.U. acknowledges financial support from the Japan Society for the Promotion of Science Overseas Fellowship. D.T. is supported by Harvard University through the Institute for Theory and Computation Fellowship. D.H. is supported by STScI grants HST-GO-17770.002, JWST-GO-12468.001, and JWST-GO-09964.001. Tomoya Kinugawa is supported by JSPS Grant-in-Aid for Transformative Research Areas (A) 23H04893 and the financial support from the Science Moves Award.
This work was made possible through the support of the Enrico Fermi Fellowships led by the Center for Spacetime and the Quantum, and supported by Grant ID \#63132 from the John Templeton Foundation. The opinions expressed in this publication are those of the author(s) and do not necessarily reflect the views of the John Templeton Foundation or those of the Center for Spacetime and the Quantum (Takatoshi Ko).

This work presents results from the European Space Agency (ESA) space mission Gaia. Gaia data are being processed by the Gaia Data Processing and Analysis Consortium (DPAC). Funding for the DPAC is provided by national institutions, in particular the institutions participating in the Gaia MultiLateral Agreement (MLA). The Gaia mission website is \url{https://www.cosmos.esa.int/gaia}. The Gaia archive website is \url{https://archives.esac.esa.int/gaia}.

The PS1 and the PS1 public science archive have been made possible through contributions by the Institute for Astronomy, the University of Hawaii, the Pan-STARRS Project Office, the Max-Planck Society and its participating institutes, the Max Planck Institute for Astronomy, Heidelberg and the Max Planck Institute for Extraterrestrial Physics, Garching, The Johns Hopkins University, Durham University, the University of Edinburgh, the Queen's University Belfast, the Harvard-Smithsonian Center for Astrophysics, the Las Cumbres Observatory Global Telescope Network Incorporated, the National Central University of Taiwan, the Space Telescope Science Institute, NASA under Grant No. NNX08AR22G issued through the Planetary Science Division of the NASA Science Mission Directorate, NSF Grant No. AST-1238877, the University of Maryland, Eotvos Lorand University, the Los Alamos National Laboratory, and the Gordon and Betty Moore Foundation.

\bibliography{manuscript}{}
\bibliographystyle{aasjournalv7}

\end{document}